\documentclass[reprint, pra, aps, amsmath, amssymb, showpacs, superscriptaddress]{revtex4-1}

\usepackage{graphicx}
\usepackage{color}
\usepackage{bm}
\usepackage{array}
\usepackage{longtable}

\begin{document}
\title{Hyperfine structure of $^2\Sigma$ molecules containing alkaline-earth atoms}

\author{Jesus Aldegunde}
\affiliation{Departamento de Quimica Fisica, Universidad de Salamanca, 37008
Salamanca, Spain}

\author{Jeremy M. Hutson}
\email{J.M.Hutson@durham.ac.uk} \affiliation{Joint Quantum Centre (JQC)
Durham-Newcastle, Department of Chemistry, Durham University, South Road,
Durham, DH1 3LE, United Kingdom}

\begin{abstract}
Ultracold molecules with both electron spin and an electric
dipole moment offer new possibilities in quantum science. We use
density-functional theory to calculate hyperfine coupling constants for a
selection of molecules important in this area, including RbSr, LiYb, RbYb, CaF
and SrF. We find substantial hyperfine coupling constants for the fermionic
isotopes of the alkaline-earth and Yb atoms. We discuss the hyperfine level
patterns and Zeeman splittings expected for these molecules. The results will
be important both to experiments aimed at forming ultracold open-shell
molecules and to their applications.
\end{abstract}

\date{\today}

\maketitle

There have recently been major advances in producing molecules in ultracold
gases of alkali-metal atoms. Ultracold molecules have been produced from most
combinations of alkali-metal atoms by magnetoassociation, in which pairs of
atoms are converted into molecules by tuning a magnetic field adiabatically
across a zero-energy Feshbach resonance. These ``Feshbach molecules" are
typically bound by less than $h\times 10$ MHz, which is less than part in
$10^7$ of the singlet well depth, and have very large internuclear separations.
A few different molecules ($^{40}$K$^{87}$Rb \cite{Ni:KRb:2008},
$^{87}$Rb$^{133}$Cs \cite{Takekoshi:RbCs:2014, Molony:RbCs:2014},
$^{23}$Na$^{40}$K \cite{Park:NaK:2015} and $^{23}$Na$^{87}$Rb
\cite{Guo:NaRb:2016}) have recently been transfered from these long-range
states to the absolute ground state by Stimulated Raman Adiabatic Passage
(STIRAP). These ground-state molecules have significant electric dipole
moments, and hold great promise for studying ultracold dipolar matter, for
precision measurement, and for applications in quantum science and technology.

The alkali-metal dimers all have singlet ground states, with no net electron
spin. This limits their tunability with magnetic fields. There is now great
interest in producing ultracold molecules with electron spin as well as an
electric dipole. Such molecules could be used to create new types of quantum
many-body systems \cite{Micheli:2006, Baranov:2012}. Promising candidates
include molecules formed from an alkali-metal atom and a laser-coolable
closed-shell atom such as Yb or Sr. \.Zuchowski {\em et al.}
\cite{Zuchowski:RbSr:2010} showed that magnetically tunable Feshbach resonances
can exist in such systems, mediated by the dependence of the alkali-metal
hyperfine coupling on the internuclear distance. Brue and Hutson
\cite{Brue:AlkYb:2013} carried out a detailed theoretical study of such
resonances in alkali metal + Yb systems. Brue and Hutson \cite{Brue:LiYb:2012}
also identified a different mechanism that can cause additional resonances in
systems containing closed-shell atoms with nuclear spin (which are all
fermionic for Sr and Yb), mediated in this case by hyperfine coupling involving
the Sr or Yb nucleus. The first Feshbach resonances of both these types have
recently been observed in RbSr \cite{Barbe:RbSr:2017}, along with resonances
due to another mechanism involving the tensorial coupling between the electron
and nuclear spins. It is likely that ultracold ground-state $^2\Sigma$
molecules of this type will be produced within the next few years.

In parallel with the work on producing ultracold molecules from atoms, there
have been major advances in direct laser-cooling of molecules such as CaF and
SrF, which also have $^2\Sigma$ ground states. Barry \emph{et al.}
\cite{Barry:2014} have cooled SrF to about 2.5~mK in a magneto-optical trap
(MOT), and Truppe \emph{et al.}\ \cite{Truppe:MOT:2017} have achieved
sub-Doppler cooling of CaF in a blue-detuned MOT to about 50~$\mu$K.

Although the basic spectroscopy of molecules in $^2\Sigma$ states is well
understood \cite{Brown:2003}, little is known quantitatively about the fine and
hyperfine coupling constants of molecules formed from alkali-metal atoms and
closed-shell atoms, or about isotopologs of CaF and SrF containing metal atoms
with non-zero spin. The magnitudes of the coupling constants will have profound
effects on the patterns of energy levels for ground-state molecules, and how
the levels cross and avoided-cross one another in magnetic, electric and laser
fields. This will in turn affect the possibilities for state transfer and
quantum control. The coupling constants are also important to understand the
strengths of Feshbach resonances \cite{Zuchowski:RbSr:2010, Brue:LiYb:2012,
Brue:AlkYb:2013, Barbe:RbSr:2017}. In this paper we present calculations of the
fine and hyperfine constants for RbSr, LiYb, RbYb, CaF and SrF, using
density-functional theory, which allow these effects to be explored.

\section{Molecular Hamiltonian}
\label{sec:ham}

The effective hamiltonian for a $^{2}\Sigma$ diatomic molecule can be written
\begin{equation}\label{hamcomp}
    H_{\rm{eff}}=H_{\rm{rfs}}+H_{\rm{hfs}}+H_{\rm{S}}+H_{\rm{Z}},
\end{equation}
where the four contributions correspond to the rotational plus fine-structure,
hyperfine, Stark and Zeeman Hamiltonians respectively.

The rotational plus fine-structure Hamiltonian $H_{\rm{rfs}}$ takes the
standard form,
\begin{equation}
    H_{\rm{rfs}}=B_v{\bm N}^2-D_v{\bm N}^2{\bm N}^2+\gamma{\bm S}\cdot{\bm N},
\label{eq:rfs}
\end{equation}
where $\bm{N}$ is the angular momentum for rotation of the molecule about its
center of mass and $\bm{S}$ is the electron spin. The third term in Eq.\
\ref{eq:rfs} represents the electron spin-rotation interaction. The hyperfine
hamiltonian $H_{\rm{hfs}}$ may be written
\begin{eqnarray}
H_{\rm{hfs}} & = &\sum_{i=1}^{2}e{\bm Q}_i\cdot{\bm q}_i
    +\sum_{i=1}^{2}{\bm S}\cdot{\bm A}_i\cdot{\bm I}_i,
\label{hameff}
\end{eqnarray}
where $\bm{I}_1$ and $\bm{I}_2$ are the spins of nuclei 1 and 2. The first term
here represents the interaction between the quadrupole tensor $e\bm{Q}_i$ of
nucleus $i$ and the electric field gradient tensor ${\bm q}_i$ at the nucleus
due to the electrons; it is commonly written in terms of a scalar nuclear
quadrupole coupling constant $(eQq)_i$. The second term represents the
interaction between the electron and nuclear spins. It is usual to separate the
isotropic and anisotropic components of the hyperfine tensor ${\bm A}_i$
\cite{kaupp:book},
\begin{equation}
    b_{\rm{F}}=A_{\rm{iso}}=\frac{A_{\|}+2A_{\bot}}{3}\mbox{ and } t
    =A_{\rm{dip}}=\frac{A_{\|}-A_{\bot}}{3},
\end{equation}
so that
\begin{equation}
{\bm S}\cdot{\bm A}_i\cdot{\bm I}_i =
    b_{{\rm F},i}{\bm S}\cdot{\bm I}_i + t_i\,\sqrt{6}\,T^2(\bm{S},\bm{I}_i)\cdot T^2(C),
    \end{equation}
where $T^2$ indicates a spherical tensor of rank 2. $T^2(C)$ has components
$C^2_q(\theta,\phi)$, where $C$ is a renormalised spherical harmonic and
$\theta,\phi$ are the polar coordinates of the internuclear vector. The
isotropic (scalar) component $b_{{\rm F},i}$ arises from the Fermi contact
interaction, whereas the anisotropic component $t_i$ arises from dipolar
interactions. The notation involving $\gamma$, $(eQq)_i$, $b_{{\rm F},i}$ and
$t_i$ coincides with that employed by Brown and Carrington \cite{Brown:2003}
(see, for example, page 607), where explicit expressions for the matrix
elements in different basis sets can be found. The alternative constants of
Frosch and Foley \cite{Frosch:pr1952} are related to these by $c_i=3t_i$ and
$b_i=b_{{\rm F},i}-t_i.$

The effect of the external fields is described by $H_{\rm{S}}$ and
$H_{\rm{Z}}$, which represent the Stark and Zeeman Hamiltonians.
The Stark Hamiltonian is
\begin{equation}
  H_{\rm{S}} = -\bm{\mu}\cdot\bm{E} - \frac{1}{2}\bm{E}\cdot\bm{\alpha}\cdot\bm{E}.
  \label{eq:Hs}
\end{equation}
It includes both a linear term to describe the interaction of the molecular
dipole $\bm{\mu}$ with a static electric field $\bm{E}$ and a quadratic term
involving the molecular polarizability tensor $\bm{\alpha}$. The latter is
usually small for static fields, but may be used with a frequency-dependent
polarizability $\bm{\alpha}(\omega)$ to account for the ac Stark effect due to
a non-resonant laser field \cite{Gregory:RbCs-AC-Stark:2017}. The Zeeman
Hamiltonian is
\begin{eqnarray}
 \nonumber H_{\rm{Z}} &=& -g_\parallel \mu_{\rm B}\bm{S}\cdot\bm{B}
 + \Delta g_{\bot} \mu_\mathrm{B} \left[ \bm{S}\cdot\bm{B}-(\bm{S}\cdot\hat{z})(\bm{B}\cdot\hat{z})\right] \\
  &&- g_{\rm r} \mu_{\rm B} \bm{N}\cdot\bm{B}
  -\sum_{i=1}^{2}g_i\mu_{\rm N}\,\bm{I}_i\cdot\bm{B} (1-\sigma_i). \label{eq:Hz}
\end{eqnarray}
The first term describes the isotropic part of the interaction of the electron
spin with an external magnetic field $\bm{B}$; $g_\parallel\approx
g_e\approx-2.0023$ is the electron g-factor parallel to the molecular axis
$\hat{z}$ and $\mu_{\rm B}$ is the Bohr magneton. The second term is an
anisotropic correction; $\Delta g_{\bot}=g_\parallel-g_\bot$, where $g_{\bot}$
is the electron $g$-factor perpendicular to the molecular axis (defined to be
negative, like $g_e$). The third and fourth terms describe the interaction of
the molecular rotation and the nuclear spins with the magnetic field; $g_{\rm
r}$ is the rotational g-factor, and $g_i$ and $\sigma_i$ are the bare nuclear
g-factor and shielding factor for nucleus $i$. $\mu_{\rm N}$ is the nuclear
magneton. The Zeeman Hamiltonian $H_{\rm{Z}}$ is dominated by the first term,
but the remaining contributions cause small shifts that may have important
consequences for resonance positions \cite{Barbe:RbSr:2017} and for the
decoherence of molecules in magnetic traps \cite{blackmore:qst2018}.

The expressions given above neglect various small terms such as the
interactions between the two nuclear spins and between the nuclear spins and
molecular rotation. These terms can be important for closed-shell molecules
\cite{Aldegunde:polar:2008, Aldegunde:nonpolar:2009, Aldegunde:spectra:2009,
Aldegunde:singlet:2017}, but for open-shell molecules they are less important
because the terms involving electron spin are always present and are two or
more orders of magnitude larger. A full description of the Hamiltonian,
including the discarded terms, can be found in Ref.~\cite{Brown:2003}.

\section{Calculation of the coupling constants}
\label{sec:calcc}

Molecular fine-structure and hyperfine constants may in principle be calculated
using either wavefunction-based methods or density-functional theory (DFT).
However, wavefunction-based methods become very complex for hyperfine
interactions in molecules containing heavy atoms, where very large basis sets
are needed and relativistic effects are important. Calculations of potential
curves for such molecules commonly use effective core potentials, but these are
of doubtful accuracy for hyperfine interactions. We therefore choose to use DFT
in the current work, and obtain values of the coupling constants $(eQq)$,
$b_{\rm{F}}$, $t$ and $\Delta g_\bot$ using the Amsterdam Density Functional (ADF)
package \cite{ADF1,ADF3}. The ADF package includes its own
all-electron basis sets of Slater functions for all the elements of the
periodic table and incorporates relativistic corrections.

In the present calculations, we employ all-electron quadruple-$\zeta$ basis
sets with four polarization functions (QZ4P). Relativistic effects are
included by means of the two-component zero-order regular approximation (ZORA)
\cite{vanLenthe:1993, vanLenthe:1994,vanLenthe:1999}. The electron
spin-rotation coupling constant, $\gamma$, is obtained from the components of
the $\bm{g}$ tensor \cite{kaupp:book} and the rotational constant using Curl's
approximation \cite{Curl:1965,bruna:pccp2003}
\begin{equation}\label{curlapp}
\gamma = -2B\,\Delta g_{\bot}
\end{equation}
According to Weltner \cite{weltner},
Curl's formula is accurate to about $\pm10\%$.

We have carried out both spin-restricted and unrestricted DFT calculations
using the B3LYP \cite{stephens:jpc1994} and PBE0 \cite{perdew:jcp1996}
functionals, for a variety of $^2\Sigma$ molecules for which experimental
values are available. The full results of these tests for the magnetic fine and
hyperfine coupling constants are given in the Supplementary Material  presented
as an appendix to the article. We conclude that spin-restricted B3LYP
calculations are the most reliable, and these results are summarized in Table
\ref{table:table1}. The largest fractional discrepancies are mostly in cases
where the constants concerned are small, and thus play a minor role for the
molecule in question. For the remaining molecules, the spin-restricted results
for $\Delta g_{\bot}$ (or equivalently $\gamma$) are accurate to 30\% or
better, with the exceptions of GaO and InO. The agreement is significantly
better for $b_{\rm{F}}$ and $t$, except for InO. The exceptions probably arise
because the ground states of these oxide radicals are mixtures of two
electronic configurations with similar energies \cite{knight:jcp1997}. Magnetic
properties are very sensitive to the balance between the configurations. The
accuracy of B3LYP calculations for nuclear quadrupole coupling constants has
been established previously \cite{Aldegunde:polar:2008, fiser:cp2013,
papai:jctc2013}.

\begin{table*}[t]
\caption{Comparison between experimental and theoretical values of $\Delta
g_{\bot}$, $\gamma$, $b_{\rm F}$ and $t$ for $^{2}\Sigma$ molecules computed
through restricted DFT calculations using the B3LYP \cite{stephens:jpc1994}
functional. An asterisk indicates cases where the signs of the components of
the ${\bm A}$ tensor were not reported in the experimental papers and have been
assigned to match the theoretical results \cite{verma:jctc2013}.  The acronyms
GP, NM and AM stand for ``Gas Phase", ``Neon Matrix" and ``Argon Matrix"
respectively and refer to the conditions used to record the spectra.
Experimental results labelled ``CA" are obtained by applying Curl's
approximation to $\Delta g_{\bot}$ or $\gamma$, depending on the case.
Theoretical values of $\gamma$ are always obtained from $\Delta g_{\bot}$ using
Curl's approximation.}
\begin{center}
\footnotesize{
\begin{tabular}{cccccccc}
  \hline\noalign{\smallskip}
  Molecule (MX) & Source & $\Delta g_{\bot}$  & $\gamma$(MHz) & $b_{\rm{F,M}}$(MHz) & $t_{\rm{M}}$(MHz) & $b_{\rm{F,X}}$(MHz) & $t_{\rm{X}}$(MHz) \\ \noalign{\smallskip}\hline\noalign{\smallskip}
   $^{103}$Rh$^{13}$C & Exp. \cite{brom:jcp1972} (NM) & 0.0518(6) & --- & $-$1097(1) & $-$8(1) & 66(1) & 11(1) \\
   & Exp. \cite{kaving:jms1969} (GP)&  --- & -1861(6) & --- & --- & --- & --- \\
   & B3LYP-R &  0.0572 & -1930 & $-$1010 & $-$2.5 & 59.3 & 8.5 \\
   \noalign{\smallskip}\hline\noalign{\smallskip}
   $^{11}$B$^{17}$O & Exp. \cite{knight:jcp1992} (NM)&  $-$0.0017(3) & $1.8(3)\times10^2$ (CA) & 1033(1) & 25(1) & $-$19(3) & $-$12(3) \\
   & B3LYP-R & $-$0.0025 & $2.61\times10^2$ & 873 & 31.1 & $-$17.0& $-$16.6  \\
   \noalign{\smallskip}\hline\noalign{\smallskip}
   $^{11}$B$^{33}$S& Exp. \cite{brom:jcp1972bs} (NM) & $-$0.0081(1) & --- & 795.6(3) & 28.9(3) & --- & --- \\
   & Exp. \cite{brom:jcp1972bs,zeeman:cjp1951} (GP) & --- & $3.8(6)\times10^2$ & --- & --- & --- & --- \\
   & B3LYP-R & $-$0.0116 & $5.46\times10^2$ & 620 & 35.3 & 13.8 & 18.7 \\
   \noalign{\smallskip}\hline\noalign{\smallskip}
   $^{27}$Al$^{17}$O & Exp. \cite{knight:jcp1997} (NM) & $-$0.0012(2) & --- & 766(1) & 52(1) & 2(1) & $-$50(1) \\
                     & Exp. \cite{yamada:jcp1990} (GP) & --- & 51.66(4) & 738(1) & 56.39(8) & --- & --- \\
   & B3LYP-R & 0.0017 & $-$62.4 & 714 & 58.1 & $-$3.9 & $-$46.4  \\
   \noalign{\smallskip}\hline\noalign{\smallskip}
   $^{69}$Ga$^{17}$O  & Exp. \cite{knight:jcp1997} (NM) & $-$0.0343(2) & 854(5) (CA) & 1483(1) & 127(1) & 8(1) & $-$77(1) \\
   & B3LYP-R & $-$0.0622 & 1550 & 1650 & 139 & 13.2 & $-$81.4  \\
   \noalign{\smallskip}\hline\noalign{\smallskip}
   $^{115}$In$^{17}$O  & Exp. \cite{knight:jcp1997} (NM) & $-$0.192(2) & $3.06(3)\times10^3$ (CA) & 1368(2) & 180(1) & 35(1) & $-$131(1)  \\
   & B3LYP-R & $-$0.337 & $5.38\times10^3$ & 2300 & 170 & 75.3 & $-$153 \\
   \noalign{\smallskip}\hline\noalign{\smallskip}
   $^{45}$Sc$^{17}$O& Exp. \cite{knight:jcp1999} (NM) & $-$0.0005(3) & 14(9) (CA) & 2018(1) & 24.7(4) & $-$20.3(3) & 0.4(2) \\
   & B3LYP-R & $-$0.0001 & 3.0 & 1850 & 13.5 & $-$22.9 & $-$0.3  \\
   \noalign{\smallskip}\hline\noalign{\smallskip}
   $^{89}$Y$^{17}$O & Exp. \cite{knight:jcp1999} (NM) & -0.0002(1) & --- & -807.5(4) & -9.5(3) & $-$16.8(2) & 0.0(2) \\
   & Exp. \cite{childs:jcp1988} (GP) & --- & $-$9.2254(1) & $-$762.976(2) & $-$9.449(1) & --- & --- \\
   & B3LYP-R & $-$0.0005 & 11.4 & $-$750 & $-$5.2 & $-$19.2 & $-$0.3  \\
   \noalign{\smallskip}\hline\noalign{\smallskip}
   $^{139}$La$^{17}$O  & Exp. \cite{knight:jcp1999} (NM) & -0.003(2) & --- & 3751(5) & 29(4) & Abs.val.$<$10 & --- \\
   & Exp. \cite{childs:jms1986} (GP) & --- & 66.1972(5) & 3631.9(1) & 31.472(1) & --- & --- \\
   & B3LYP-R & $-$0.0046 & 91.3 & 3460 & 16.6 & $-$12.5 & $-$0.6 \\
   \noalign{\smallskip}\hline\noalign{\smallskip}
   $^{67}$Zn$^{1}$H & Exp. \cite{mckinley:jpca2000} (NM) & $-$0.0182(3) & $7.2(1)\times10^3$ (CA) & 630(1) & 15(1) & 503(1) & $-$1(1) \\
   & B3LYP-R & $-$0.0244 & $9.79\times10^3$ & 616 & 23.8 & 382 & 1.4  \\
   \noalign{\smallskip}\hline\noalign{\smallskip}
   $^{67}$Zn$^{19}$F& Exp. \cite{knight:jmr1978} (NM) & $-$0.006(1) & $1.3(2)\times10^2$ (CA) & --- & --- & 319(2) & 177(2) \\
   & B3LYP-R & $-$0.0073 & $1.59\times10^2$ &  1160 & 15.4 & 266 & 210  \\
    \noalign{\smallskip}\hline\noalign{\smallskip}
   $^{111}$Cd$^{19}$F & Exp. \cite{knight:jmr1978} (NM) & $-$0.017(2) & $4.8(6)\times10^2$ (CA) & --- & --- & 266(3) & 202(2) \\
   & B3LYP-R & $-$0.0314 & $8.79\times10^2$ & $-$3600 & $-$255 & 567 & 229   \\
   \noalign{\smallskip}\hline\noalign{\smallskip}
   $^{67}$Zn$^{107}$Ag & Exp. \cite{kasai:jpc1978} (AM) & $-$0.0118(2) & $39(1)$ (CA) & --- & --- & $-$1324(3)$^{*}$ & 0(1) \\
   & B3LYP-R & $-$0.0158 & 52.0 & 306 & 6.9 & $-$1250 & $-$0.6  \\
  \noalign{\smallskip}\hline\noalign{\smallskip}
   $^{105}$Pd$^{1}$H & Exp. \cite{knight:jcp1990} (AM) & 0.291(1) & $-1.252(4)\times10^5$ (CA) & $-$823(4) & $-$22(3) & --- & --- \\
   & Exp. \cite{knight:jcp1990} (NM) & 0.291(1) & $-1.252(4)\times10^5$ (CA) & $-$857(4) & $-$16(3) & --- & --- \\
   & B3LYP-R & 0.266 &  $-1.14\times10^5$ & $-$914 & $-$2.4 & 117 & 7.0  \\
  \noalign{\smallskip}\hline\noalign{\smallskip}
   $^{111}$Cd$^{1}$H & Exp. \cite{tan:jcp1994} (GP) & $-$0.0567(2) (CA) & $1.811(6)\times10^4$ & $-$3764(26) & $-$122(6) & 558(10) & --- \\
   & Exp. \cite{varberg:jms2004} (GP) & --- & --- & $-$3766.3(15) & $-$143(1) & 549.8(18) & $-$2.4(8) \\
   & B3LYP-R & $-$0.0735 & $2.4\times10^4$ & $-$3920 & $-$175 & 374 & 0.9  \\
  \noalign{\smallskip}\hline\noalign{\smallskip}
  $^{111}$Cd$^{107}$Ag & Exp. \cite{kasai:jpc1978} (AM) & $-$0.0312(2) &68.9(4) & $-$2053(3)$^{*}$ & $-$63(3)$^{*}$ & $-$1327(3)$^{*}$ & 0(1) \\
   & B3LYP-R & $-$0.0400&  88.4 & $-$2010 &  $-$55.4 & $-$1210 & $-$0.6  \\
  \noalign{\smallskip}\hline\noalign{\smallskip}
   $^7$Li$^{40}$Ca & Exp. \cite{ivanova:jcp2011} (GP) & $-$0.0068(1) (CA) & 103(2) & --- & --- & --- & ---  \\
   & B3LYP-R & $-$0.0119 & 179 & 218 & 0.2 & $-$107 & $-$4.6 \\
   \noalign{\smallskip}\hline\noalign{\smallskip}
   $^7$Li$^{138}$Ba & Exp. \cite{dincan:jcp1994} (GP) & $-$0.1205(1) (CA) & 1384.5(9) & --- & --- & --- & ---  \\
   & B3LYP-R & $-$0.129 &  1480 & 162 &  0.3 &  806 &   28.1  \\
   \noalign{\smallskip}\hline\noalign{\smallskip}
   $^{40}$Ca$^{19}$F & Exp. \cite{childs:jms1981} (GP) & $-$0.00193(1) (CA) & 39.49793(2)  & --- & --- &  122.025(1) & 13.549(1)  \\
   & B3LYP-R & $-$0.00180 & 37.2 & --- & --- & 127 & 8.0 \\
   \noalign{\smallskip}\hline\noalign{\smallskip}
   $^{88}$Sr$^{19}$F & Exp. \cite{childs:jms1981b} (GP) & -0.00495(1) (CA) & 74.79485(10) & --- & --- & 107.1724(10) & 10.089(10)  \\
   & B3LYP-R & $-$0.00463 & 69.9 & --- & --- & 112 & 6.8 \\
   \noalign{\smallskip}\hline\noalign{\smallskip}
\end{tabular}}
\end{center}
\label{table:table1} 
\end{table*}

The ADF program produces values of the coupling constants for a single
isotopolog, usually the one containing the most abundant isotopes. Coupling
constants for other isotopologs are obtained using simple scalings involving
rotational constants, nuclear $g$-factors and nuclear quadrupole moments.

\section{Coupling constants for R\MakeLowercase{b}S\MakeLowercase{r}, L\MakeLowercase{i}Y\MakeLowercase{b},
R\MakeLowercase{b}Y\MakeLowercase{b},
C\MakeLowercase{a}F and S\MakeLowercase{r}F} \label{sec:cc}

{Table \ref{table:table2} gives the coupling constants for all stable
isotopologs of RbSr, LiYb, RbYb, CaF and SrF, obtained from spin-restricted
calculations at the equilibrium geometries, $R_{\rm{e}}$=4.67 \AA\ for RbSr
\cite{zuchowski:pra2014}, 3.52 \AA\ for LiYb \cite{Brue:AlkYb:2013}, 4.91 \AA\
for RbYb \cite{Brue:AlkYb:2013}, 1.95 \AA\ for CaF \cite{childs:jms1981} and
2.07 \AA\ for SrF \cite{colarusso:jms1996}. The spin-restricted results for one
isotopolog of each molecule are compared with unrestricted results in Table
\ref{table:table3}; the differences are mostly within 20\%, although for LiYb
some of them approach 30\%.

Experimental results are available for CaF \cite{childs:jms1981} and SrF
\cite{childs:jms1981b}, but only for isotopologs containing metal atoms with
zero nuclear spin. The agreement between the experimental and theoretical
results is good, with errors below 15\% for CaF and SrF. The present results
also agree with previous calculations of $b_{\rm F}$ as a function of
internuclear distance for Rb in RbSr \cite{Zuchowski:RbSr:2010} and Rb in RbYb
\cite{Brue:AlkYb:2013}.

In molecular spectroscopy, a $^2\Sigma$ molecule without nuclear spin is
commonly described using Hund's case (B), in which the electron spin $S$
couples to the molecular rotation $N$ to form a resultant $J$. However, $J$ is
a useful quantum number only if the hyperfine interactions are weak compared to
the spin-rotation interaction, which is not the case for most of the molecules
considered here. In the present work we couple the electron and
nuclear spins \emph{before} coupling their vector sum to the molecular
rotation.

There is some difficulty in choosing a notation for molecular quantum numbers
that does not clash with usage in either atomic physics or molecular
spectroscopy. In molecular spectroscopy, $F$ is commonly used for the total
angular momentum of a molecule, including rotation and all spins. However, in
atomic physics, $F$ is often used for the total angular momentum of a single
atom. For collision problems and Van der Waals complexes, there is a
well-established convention that quantum numbers that apply to individual
colliding species (or monomers) are converted to lower-case, reserving the
upper-case letter for the corresponding quantum number of the collision complex
\cite{Dubernet:1993}. We follow this convention here and retain $s$, $i$ and
$f$ for the electron spin, nuclear spin and total angular momentum of
individual atoms, and use $F$ for the resultant of $f_1$ and $f_2$. In our
notation, $F$ is thus the total angular momentum of the molecule
\emph{excluding rotation}. This accords with usage in systems such as RbCs and
Cs$_2$ \cite{Takekoshi:RbCs:2012, Berninger:Cs2:2013}, although Brown and
Carrington \cite{Brown:2003} use $G$ in this context. We use $N$ for the
mechanical rotation of the pair (equivalent to the partial-wave quantum number
$L$ in collisions). We designate the total angular momentum of the molecule
$\cal F$, the resultant of $F$ and $N$.  All the quantum numbers can have
projections denoted $m_i$, $M_F$, etc., which may be nearly conserved in
certain field regimes.

Figure \ref{fig:01} shows the Zeeman splitting of the hyperfine levels for the
lowest two rotational levels of $^{87}$Rb$^{88}$Sr at magnetic fields up to
20~G. The hyperfine coupling constant $b_{\rm F,Rb}$ is 2.60~GHz, which is
reduced by about 25\% from its atomic value of 3.42~GHz. The resulting
splitting is 5.2~GHz, which is considerably larger than the rotational spacing
of 1.1~GHz, so levels correlating with $f=2$ are well off the top of Fig.\
\ref{fig:01}. The rotationless $N=0$ state, with $f=1$ and $F=1$, splits into 3
sublevels with projection $M_F$, just like a free $^{87}$Rb atom. By contrast,
the $N=1$ state with $F=f=1$ is split into three zero-field levels with ${\cal
F}=0$, 1 and 2 by the spin-rotation coupling. When a magnetic field is applied,
each of these splits initially into $2{\cal F}+1$ components labeled by the
total projection $M_{\cal F}$. However, states of the same $M_{\cal F}$
originating from different ${\cal F}$ levels mix as the field increases; at
higher fields, $\cal F$ is no longer a good quantum number and the magnetic
sublevels are better described by $M_F$ and $M_N$. In this regime, from 30 to
about 1000~G, $F=f$ remains nearly conserved. At even higher fields, levels of
different $F$ will mix and eventually the best quantum numbers are $M_S=m_s$,
$M_I=m_i$ and $M_N$.

The situation is more complicated when the closed-shell atom has non-zero
nuclear spin. We consider briefly the example of $^{87}$Rb$^{87}$Sr, which is
topical because Feshbach resonances have recently been observed for this
combination \cite{Barbe:RbSr:2017}. The largest coupling is still between $S$
and $i_{\rm Rb}$ to form $f_{\rm Rb}=1$ and 2, but in this case $f_{\rm Rb}=1$
couples to $i_{\rm Sr}=9/2$ to form $F=7/2$, 9/2 and 11/2. There are thus 3
zero-field states even for $N=0$, spread over about 160~MHz by the coupling
between $i_{\rm Sr}$ and $S$. For $N=1$, these are each split into 3 by the
spin-rotation coupling: $F=7/2\rightarrow {\cal F}=5/2,7/2,9/2$,
$F=9/2\rightarrow {\cal F}=7/2,9/2,11/2$, $F=11/2\rightarrow {\cal
F}=9/2,11/2,13/2$. In a magnetic field these split into a total of $(2f_{\rm
Rb}+1)(2i_{\rm Sr}+1)(2N+1)=90$ sublevels. The different angular momenta
decouple sequentially as the magnetic field increases: first $N$, then $i_{\rm
Sr}$ and finally $i_{\rm Rb}$. For $N>0$ there are additional hyperfine
couplings due to nuclear quadrupole interactions [$(eQq)_{\rm Rb}$ and
$(eQq)_{\rm Sr}$] and anisotropic electron-nuclear spin couplings ($t_{\rm Rb}$
and $t_{\rm Sr}$); these shift the resulting levels by a few MHz, but do not
produce additional splittings. The resulting Zeeman diagram is very complicated
and is beyond the scope of this paper to explore in detail.

The situation is different again for CaF and SrF. Here the chemical interaction
is strong enough that an atomic $f$ quantum number for fluorine is not useful.
The coupling between the electron and nuclear spins is much \emph{smaller} than
the separation between molecular rotational levels, so the ordering of levels
is different. For even-mass Ca or Sr isotopes with $i=0$ the primary coupling
is between $S=1/2$ and $i_{\rm F}=1/2$ to form $F=0$ and 1. The resulting
levels have been explored in previous work \cite{childs:jms1981}. For $^{43}$Ca
and $^{87}$Sr, however, the primary coupling is between $S=1/2$ and $i_{\rm
Ca}=7/2$ or $i_{\rm Sr}=9/2$. For $^{87}$SrF these couple to form levels with
$f_{\rm Sr}=4$ and $5$, separated by about 2.6~GHz. These levels are then
further split by weaker coupling to $i_{\rm F}=1/2$ to form zero-field $N=0$
states $F=7/2$, 9/2, 9/2 and 11/2. For $N>0$ these are further split by
spin-rotation coupling. $^{43}$CaF behaves analogously.

It is noteworthy that both the isotropic and dipolar magnetic hyperfine
couplings are a factor of 7 to 10 stronger for $^{171}$Yb in RbYb than for
$^{87}$Sr in RbSr. This makes $^{171}$Yb a particularly appealing candidate for
Feshbach resonances such as those predicted in ref.\ \cite{Brue:LiYb:2012} and
observed for $^{87}$Rb$^{87}$Sr in ref.\ \cite{Barbe:RbSr:2017}.

\section{Conclusions} \label{sec:conc}

Hyperfine coupling in $^2\Sigma$ molecules containing alkaline-earth atoms is
important both in producing ultracold molecules and in using them for
applications in quantum science. We have used density-functional theory to
calculate hyperfine coupling constants for several $^2\Sigma$ molecules that
are the targets of current experiments aimed at producing ultracold molecules.
We have focused on molecules formed from an alkaline-earth (or Yb) atom and
either an alkali-metal atom or fluorine. The resulting hyperfine splitting
patterns and Zeeman splittings are illustrated by considering isotopologs of
RbSr and SrF.

%
%
\begin{figure}
  \includegraphics[width=0.44\textwidth]{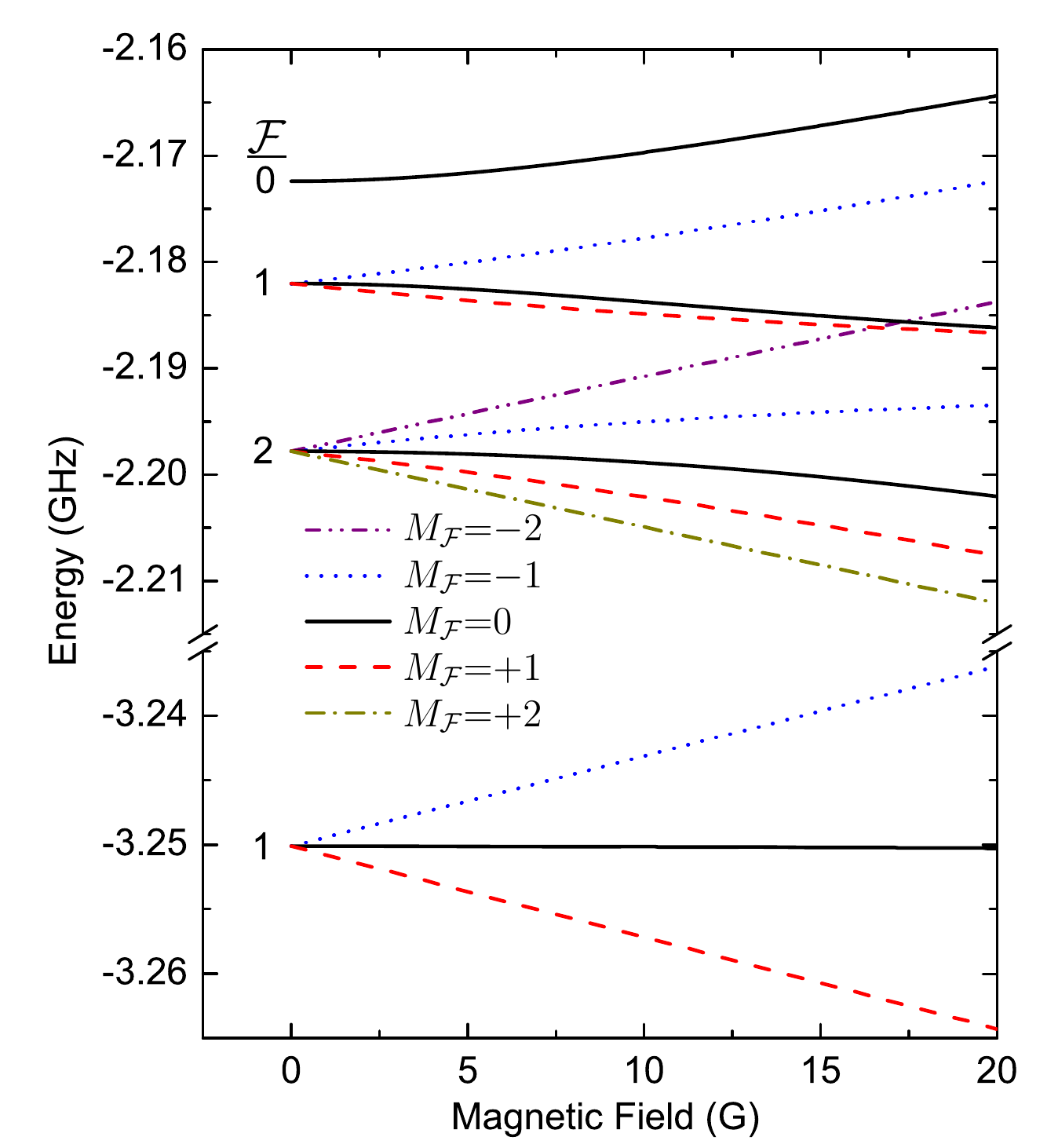}
  \caption{\label{fig:01}%
    Zeeman splitting of the lowest hyperfine energy levels of $^{87}{\rm Rb}^{88}{\rm Sr}$
    for magnetic fields up to 20 G.}
\end{figure}

\begin{table*}
\caption{Coupling constants for the isotopologs of RbSr, LiYb, RbYb, CaF
and SrF.} 
\begin{tabular}{ccccccccccc}
  \hline\noalign{\smallskip}
  AX & $I_{\rm{A}}$ & $I_{\rm{X}}$ & $\gamma$/MHz & $b_{\rm{F,A}}$(MHz) & $t_{\rm{A}}$(MHz) & $b_{\rm{F,X}}$(MHz) & $t_{\rm{X}}$(MHz) & $(eQq)_{\rm{A}}$/MHz & $(eQq)_{\rm{X}}$/MHz \\ \noalign{\smallskip}\hline\noalign{\smallskip}
  $^{85}$Rb$^{84}$Sr &  5/2 & 0 & 33.8 & 767 & 0.01 & --- & ---& $-$7.5 & --- \\
  $^{85}$Rb$^{86}$Sr &  5/2 & 0 & 33.4 & 767 & 0.01 & --- & --- & $-$7.5 & --- \\
  $^{85}$Rb$^{87}$Sr &  5/2 & 9/2 & 33.2 & 767 & 0.01 & $-$65.2 & $-$3.7 & $-$7.5 & $-$23.1 \\
  $^{85}$Rb$^{88}$Sr &  5/2 & 0 & 33.0 & 767 & 0.01 & --- & --- & $-$7.5 & --- \\
  $^{87}$Rb$^{84}$Sr &  3/2 & 0 & 33.4 & 2600 & 0.04 & --- & --- & $-$3.6 & --- \\
  $^{87}$Rb$^{86}$Sr &  3/2 & 0 & 33.0 & 2600 & 0.04 & --- & --- & $-$3.6 & --- \\
  $^{87}$Rb$^{87}$Sr &  3/2 & 9/2 & 32.8 & 2600 & 0.04 & $-$65.2 & $-$3.7 & $-$3.6 & $-$23.1 \\
  $^{87}$Rb$^{88}$Sr &  3/2 & 0 & 32.6 & 2600 & 0.04 & --- & --- & $-$3.6 & --- \\
  \noalign{\smallskip}\hline\noalign{\smallskip}
  $^{6}$Li$^{168}$Yb &  1 & 0 &  1880 & 97.2 & 0.1 & --- & --- & $\approx$0 & --- \\
  $^{6}$Li$^{170}$Yb &  1 & 0 & 1880 & 97.2 & 0.1 & --- & ---& $\approx$0 & --- \\
  $^{6}$Li$^{171}$Yb &  1 & 1/2 & 1880 & 97.2 & 0.1 & 1440 & 83.1 & $\approx$0& --- \\
  $^{6}$Li$^{172}$Yb &  1 & 0 & 1880 & 97.2 & 0.1 & --- & --- & $\approx$0& --- \\
  $^{6}$Li$^{173}$Yb &  1 & 5/2 & 1880 & 97.2 & 0.1 & $-$396 & $-$22.9 & $\approx$0 & $-$786 \\
  $^{6}$Li$^{174}$Yb &  1 & 0 &  1880 & 97.2 & 0.1 & --- & --- & $\approx$0 & --- \\
  $^{6}$Li$^{176}$Yb &  1 & 0 &  1880 & 97.2 & 0.1 & --- & --- & $\approx$0 & --- \\
  $^{7}$Li$^{168}$Yb &  3/2 & 0 &  1620 & 257 & 0.2 & --- & --- & 0.1 & --- \\
  $^{7}$Li$^{170}$Yb &  3/2 & 0 &  1620 & 257 & 0.2 & --- & --- & 0.1 & --- \\
  $^{7}$Li$^{171}$Yb &  3/2 & 1/2 & 1620 & 257 & 0.2 & 1440 & 83.1 & 0.1 & --- \\
  $^{7}$Li$^{172}$Yb &  3/2 & 0 &  1620 & 257 & 0.2 & --- & --- & 0.1 & --- \\
  $^{7}$Li$^{173}$Yb &  3/2 & 5/2 & 1620 & 257 & 0.2 & $-$396 & $-$22.9 & 0.1 & $-$786 \\
  $^{7}$Li$^{174}$Yb &  3/2 & 0 &  1620 & 257 & 0.2 & --- & --- & 0.1 & --- \\
  $^{7}$Li$^{176}$Yb &  3/2 & 0 &  1620 & 257 & 0.2 & --- & --- & 0.1 & --- \\ \noalign{\smallskip}\hline\noalign{\smallskip}
  $^{85}$Rb$^{168}$Yb &  5/2 & 0 & 54.9 & 844 & 0.02 & --- & --- & $-$5.1 & --- \\
  $^{85}$Rb$^{170}$Yb &  5/2 & 0 & 54.7 &  844 & 0.02 & --- & --- & $-$5.1 & --- \\
  $^{85}$Rb$^{171}$Yb &  5/2 & 1/2 & 54.6 & 844 & 0.02 & 499 & 36.8 & $-$5.1 & --- \\
  $^{85}$Rb$^{172}$Yb &  5/2 & 0 & 54.5 & 844 & 0.02 & --- & --- & $-$5.1 & --- \\
  $^{85}$Rb$^{173}$Yb &  5/2 & 5/2 & 54.4 & 844 & 0.02 & $-$137 & $-$10.1 & $-$5.1 & $-$303 \\
  $^{85}$Rb$^{174}$Yb &  5/2 & 0 & 54.3 & 844 & 0.02 & --- & --- & $-$5.1 & --- \\
  $^{85}$Rb$^{176}$Yb &  5/2 & 0 & 54.1 & 844 & 0.02 & --- & --- & $-$5.1 & --- \\
  $^{87}$Rb$^{168}$Yb &  3/2 & 0 & 54.1 & 2860 & 0.1 & --- & --- & $-$2.3 & --- \\
  $^{87}$Rb$^{170}$Yb &  3/2 & 0 & 53.9 & 2860 & 0.1 & --- & --- & $-$2.3 & --- \\
  $^{87}$Rb$^{171}$Yb &  3/2 & 1/2 & 53.7 & 2860 & 0.1 & 499 & 36.8 & $-$2.3 & --- \\
  $^{87}$Rb$^{172}$Yb &  3/2 & 0 & 53.6 & 2860 & 0.1 & --- & --- & $-$2.3 & --- \\
  $^{87}$Rb$^{173}$Yb &  3/2 & 5/2 & 53.5 & 2860 & 0.1 & $-$137 & $-$10.1 & $-$2.3 & $-$303 \\
  $^{87}$Rb$^{174}$Yb &  3/2 & 0 & 53.4 & 2860 & 0.1 & --- & --- & $-$2.3 & --- \\
  $^{87}$Rb$^{176}$Yb &  3/2 & 0 & 53.2 & 2860 & 0.1 & --- & --- & $-$2.3 & --- \\
  \noalign{\smallskip}\hline\noalign{\smallskip}
  $^{40}$Ca$^{19}$F &  0 & 1/2 & 37.2 & --- & --- & 127 & 8.0 & --- & --- \\
  $^{42}$Ca$^{19}$F &  0 & 1/2 & 36.6  & --- & --- & 127 & 8.0 & --- & --- \\
  $^{43}$Ca$^{19}$F &  7/2 & 1/2 & 36.3  & $-$404 & $-$3.0 & 127 & 8.0 & 9.9 & --- \\
  $^{44}$Ca$^{19}$F &  0 & 1/2 & 36.1  & --- & --- & 127 & 8.0 & --- & --- \\
  $^{46}$Ca$^{19}$F &  0 & 1/2 & 35.6  & --- & --- & 127 & 8.0 & --- & --- \\
  $^{48}$Ca$^{19}$F &  0 & 1/2 & 35.2  & --- & --- & 127 & 8.0 & --- & --- \\
  \noalign{\smallskip}\hline\noalign{\smallskip}
  $^{84}$Sr$^{19}$F &  0 & 1/2 & 70.5 & --- & --- & 112 & 6.8 & --- & --- \\
  $^{86}$Sr$^{19}$F &  0 & 1/2 & 70.2 & --- & --- & 112 & 6.8 & --- & --- \\
  $^{87}$Sr$^{19}$F &  9/2 & 1/2 & 70.1 & $-$525 & $-$3.8 & 112 & 6.8 & $-$150 & --- \\
  $^{88}$Sr$^{19}$F &  0 & 1/2 & 69.9  & --- & --- &  112 & 6.8 & --- & --- \\
  \noalign{\smallskip}\hline
\end{tabular}
\label{table:table2} 
\end{table*}

\begin{table*}
\caption{Coupling constants for one isotopolog of RbSr, LiYb, RbYb, CaF and SrF calculated using restricted and unrestricted calculations.} 
\begin{tabular}{ccccccccccc}
  \hline\noalign{\smallskip}
  AX & $I_{\rm{A}}$ & Source & $I_{\rm{X}}$ & $\gamma$/MHz & $b_{\rm{F,A}}$(MHz) & $t_{\rm{A}}$(MHz) & $b_{\rm{F,X}}$(MHz) & $t_{\rm{X}}$(MHz) & $(eQq)_{\rm{A}}$/MHz & $(eQq)_{\rm{X}}$/MHz \\ \noalign{\smallskip}\hline\noalign{\smallskip}
  $^{85}$Rb$^{87}$Sr &  B3LYP-R & 5/2 & 9/2 & 33.2 & 767 & 0.01 & $-$65.2 & $-$3.7 & $-$7.5 & $-$23.1 \\
                     &  B3LYP-U & 5/2 & 9/2 & 26.7 & 893 & $-$0.50 & $-$52.1  & $-$3.8 & $-$7.3 & $-$23.3 \\
  \noalign{\smallskip}\hline\noalign{\smallskip}
  $^{7}$Li$^{173}$Yb &  B3LYP-R & 3/2 & 5/2 & 1620 & 257 & 0.2 & $-$396 & $-$22.9 & 0.1 & $-$786 \\
                     &  B3LYP-U & 3/2 & 5/2 & 1190 & 364 & 0.2 & $-$294 & $-$20.5 & 0.1 & $-$673 \\
  \noalign{\smallskip}\hline\noalign{\smallskip}
  $^{85}$Rb$^{173}$Yb &  B3LYP-R & 5/2 & 5/2 & 54.4 & 844 & 0.02 & $-$137 & $-$10.1 & $-$5.1 & $-$303 \\
                      &  B3LYP-U & 5/2 & 5/2 & 40.9 & 962 & $-$0.16 & $-$105  & $-$9.0 & $-$4.8 & $-$257 \\
  \noalign{\smallskip}\hline\noalign{\smallskip}
  $^{43}$Ca$^{19}$F &  B3LYP-R & 7/2 & 1/2 & 36.3  & $-$404 & $-$3.0 & 127 & 8.0 & 9.9 & --- \\
                    &  B3LYP-U & 7/2 & 1/2 & 38.9  & $-$443 & $-$4.8 & 126 & 8.2 & 10.2 & --- \\
  \noalign{\smallskip}\hline\noalign{\smallskip}
  $^{87}$Sr$^{19}$F &  B3LYP-R & 9/2 & 1/2 & 70.1 & $-$525 & $-$3.8 & 112 & 6.8 & $-$150 & --- \\
                    &  B3LYP-U & 9/2 & 1/2 & 65.2 & $-$570 & $-$6.0 & 114 & 7.0 & $-$154 & --- \\
  \noalign{\smallskip}\hline
\end{tabular}
\label{table:table3} 
\end{table*}

\appendix*

\section{Supplemental Material}

Table \ref{table:table4} gives results for both spin-restricted and spin-unrestricted DFT
calculations using the B3LYP \cite{stephens:jpc1994} and PBE0
\cite{perdew:jcp1996} functionals, for a variety of $^2\Sigma$ molecules for
which experimental values are available. Overall, B3LYP is a little more
accurate than PBEO and so we use B3LYP in the main paper. The largest fractional
discrepancies are mostly in cases where the constants concerned are small, and
thus play a minor role for the molecule in question. In these cases the
calculations correctly give small values, though sometimes with substantial
percentage errors. For the remaining molecules, the spin-restricted results for
$\Delta g_{\bot}$ (or equivalently $\gamma$) are accurate to 30\% or better,
with the exceptions of GaO and InO. The agreement is significantly better for
$b_{\rm{F}}$ and $t$, except for InO. The exceptions probably arise because the
ground states of these oxide radicals are mixtures of two electronic
configurations with similar energies \cite{knight:jcp1997}. Molecular
properties such as hyperfine coupling constants are very sensitive to the
balance between the configurations.

Unrestricted calculations are often slightly more accurate than restricted
calculations, especially for $\gamma$. However, in some cases they give very
poor results, even where the fine and hyperfine coupling constants are large:
see, for example, the values of $b_{\rm{F}}$ for the metals in AlO, GaO and
InO. It appears that unrestricted calculations on these oxides are even more
susceptible to mixing of configurations than restricted calculations. The
unrestricted B3LYP calculation also give dramatically incorrect results for
$\gamma$ in LiBa: in this case we have calculated the effective spin of the
molecule, and find that its value is far from 1/2 in the unrestricted case, so
it is clear that the solution suffers from spin contamination.

We conclude that spin-restricted B3LYP calculations give the most reliable
overall results. It is however valuable carry out unrestricted calculations as
well: in cases where the two are similar, the unrestricted result may be
better.

\begin{longtable*}{cccccccc}
\caption{Comparison between experimental and theoretical
values of $\Delta g_{\bot}$, $\gamma$, $b_{\rm F}$ and $t$ for $^{2}\Sigma$
molecules computed through restricted (R) and unrestricted (U) DFT calculations
using the B3LYP \cite{stephens:jpc1994} and PBE0 \cite{perdew:jcp1996}
functionals. An asterisk indicates cases where the signs of the components of the ${\bm A}$
tensor were not reported in the experimental papers and have been assigned to
match the theoretical results \cite{verma:jctc2013}.  The acronyms GP, NM and
AM stand for ``Gas Phase", ``Neon Matrix" and ``Argon Matrix" respectively and
refer to the conditions used to record the spectra. Experimental results
labelled as ``CA" are obtained by applying Curl's approximation to $\Delta
g_{\bot}$ or $\gamma$, depending on the case. Theoretical values of $\gamma$
are always obtained from $\Delta g_{\bot}$ using Curl's approximation.} \label{table:table4}\\%
\hline
Molecule (MX) & Source & $\Delta g_{\bot}$  & $\gamma$(MHz) & $b_{\rm{F,M}}$(MHz) & $t_{\rm{M}}$(MHz) & $b_{\rm{F,X}}$(MHz) & $t_{\rm{X}}$(MHz) \\
\endfirsthead
\hline
Molecule (MX) & Source & $\Delta g_{\bot}$  & $\gamma$(MHz) & $b_{\rm{F,M}}$(MHz) & $t_{\rm{M}}$(MHz) & $b_{\rm{F,X}}$(MHz) & $t_{\rm{X}}$(MHz) \\
\hline
\endhead
\hline \multicolumn{8}{r}{\textit{Continued on next page}} \\
\endfoot
\endlastfoot
\noalign{\smallskip}\hline\noalign{\smallskip}
   $^{103}$Rh$^{13}$C & Exp. \cite{brom:jcp1972} (NM) & 0.0518(6) & --- & $-$1097(1) & $-$8(1) & 66(1) & 11(1) \\
   & Exp. \cite{kaving:jms1969} (GP)&  --- & -1861(6) & --- & --- & --- & --- \\
   & B3LYP-U &  0.0720  & $-$2420 & $-$1080 & $-$6.7 & 60.0 & 13.5 \\
   & B3LYP-R &  0.0572 & $-$1930 & $-$1010 & $-$2.5 & 59.3 & 8.5 \\
   & PBE0-U &  0.0799 & $-$2690 & $-$1080 & $-$7.9 & 46.3 & 13.8 \\
   & PBE0-R &  0.0625 & $-$2100 & $-$999 & $-$2.8 & 55.2 & 8.4  \\
   \noalign{\smallskip}\hline\noalign{\smallskip}
   $^{11}$B$^{17}$O & Exp. \cite{knight:jcp1992} (NM)&  $-$0.0017(3) & $1.8(3)\times10^2$ (CA) & 1033(1) & 25(1) & $-$19(3) & $-$12(3) \\
   & B3LYP-U & $-$0.0023 & $2.38\times10^2$ & 1080 & 29.1 & $-$10.7& $-$21.5  \\
   & B3LYP-R & $-$0.0025 & $2.61\times10^2$ & 873 & 31.1 & $-$17.0& $-$16.6  \\
   & PBE0-U & $-$0.0023 & $2.43\times10^2$ & 1040 & 26.9 & $-$10.4& $-$23.3  \\
   & PBE0-R & $-$0.0024 & $2.55\times10^2$ & 829  & 30.2 & $-$17.8& $-$16.6  \\
   \noalign{\smallskip}\hline\noalign{\smallskip}
   $^{11}$B$^{33}$S& Exp. \cite{brom:jcp1972bs} (NM) & $-$0.0081(1) & --- & 795.6(3) & 28.9(3) & --- & --- \\
   & Exp. \cite{brom:jcp1972bs,zeeman:cjp1951} (GP) & --- & $3.8(6)\times10^2$ & --- & --- & --- & --- \\
   & B3LYP-U & $-$0.0102 & $4.80\times10^2$ & 824 & 34.0 & 2.3 & 22.1 \\
   & B3LYP-R & $-$0.0116 & $5.46\times10^2$ & 620 & 35.3 & 13.8 & 18.7 \\
   & PBE0-U & $-$0.0101 & $4.78\times10^2$ & 805 & 31.4 & 3.4 & 23.3 \\
   & PBE0-R & $-$0.0108 & $5.12\times10^2$ & 595 & 33.8 & 14.4 & 18.8 \\
   \noalign{\smallskip}\hline\noalign{\smallskip}
   $^{27}$Al$^{17}$O & Exp. \cite{knight:jcp1997} (NM) & $-$0.0012(2) & --- & 766(1) & 52(1) & 2(1) & $-$50(1) \\
                     & Exp. \cite{yamada:jcp1990} (GP) & --- & 51.66(4) & 738(1) & 56.39(8) & --- & --- \\
   & B3LYP-U & 0.0007 & $-$26.6 & 472 & 62.2 & 7.6 & $-$64.8  \\
   & B3LYP-R & 0.0017 & $-$62.4 & 714 & 58.1 & $-$3.9 & $-$46.4  \\
   & PBE0-U & $-$0.0002 & 6.7 & 434 & 60.3 & 18.6 & $-$61.9  \\
   & PBE0-R & 0.0010 & $-$35.1 & 687 & 56.1 &  $-$3.5 & $-$43.9 \\
   \noalign{\smallskip}\hline\noalign{\smallskip}
   $^{69}$Ga$^{17}$O  & Exp. \cite{knight:jcp1997} (NM) & $-$0.0343(2) & 854(5) (CA) & 1483(1) & 127(1) & 8(1) & $-$77(1) \\
   & B3LYP-U & $-$0.0387 & 965 & 635 & 142 & 12.3 & $-$95.8  \\
   & B3LYP-R & $-$0.0622 & 1550 & 1650 & 139 & 13.2 & $-$81.4  \\
   & PBE0-U & $-$0.0354 & 883 & 536 & 142 & 25.3 & $-$93.1 \\
   & PBE0-R & $-$0.0561 & 1400 & 1670 & 139 & 11.1 & $-$75.5  \\
   \noalign{\smallskip}\hline\noalign{\smallskip}
   $^{115}$In$^{17}$O  & Exp. \cite{knight:jcp1997} (NM) & $-$0.192(2) & $3.06(3)\times10^3$ (CA) & 1368(2) & 180(1) & 35(1) & -131(1)  \\
   & B3LYP-U & $-$0.152 & $2.42\times10^3$ & 389 & 221 & 27.9 & $-$125 \\
   & B3LYP-R & $-$0.337 & $5.38\times10^3$ & 2300 & 170 & 75.3 & $-$153 \\
   & PBE0-U & $-$0.137 & $2.15\times10^3$ & 205 & 232 & 37.7 & $-$120 \\
   & PBE0-R & $-$0.270 & $4.26\times10^3$ &  2390 &  194 & 59.5 & $-$130  \\
   \noalign{\smallskip}\hline\noalign{\smallskip}
   $^{45}$Sc$^{17}$O& Exp. \cite{knight:jcp1999} (NM) & $-$0.0005(3) & 14(9) (CA) & 2018(1) & 24.7(4) & $-$20.3(3) & 0.4(2) \\
   & B3LYP-U & $-$0.0007 & 20.9 & 1990 & 22.1 & $-$20.2 & 0.7  \\
   & B3LYP-R & $-$0.0001 & 3.0 & 1850 & 13.5 & $-$22.9 & -0.3  \\
   & PBE0-U & $-$0.0012 & 34.6 & 1830 & 22.1 & $-$16.1 & 0.5  \\
   & PBE0-R & $-$0.0003 & 10.2 & 1690 & 13.1 & $-$21.1 & -0.3 \\
   \noalign{\smallskip}\hline\noalign{\smallskip}
   $^{89}$Y$^{17}$O & Exp. \cite{knight:jcp1999} (NM) & -0.0002(1) & --- & -807.5(4) & -9.5(3) & $-$16.8(2) & 0.0(2) \\
   & Exp. \cite{childs:jcp1988} (GP) & --- & $-$9.2254(1) & $-$762.976(2) & $-$9.449(1) & --- & --- \\
   & B3LYP-U & $-$0.0004 & 8.7 & $-$804 & $-$8.0 & $-$17.7 & 0.3 \\
   & B3LYP-R & $-$0.0005 & 11.4 & $-$750 & $-$5.2 & $-$19.2 & $-$0.3  \\
   & PBE0-U & $-$0.0013 & 28.7 & $-$749 & $-$8.1 & $-$13.8 & 0.3  \\
   & PBE0-R & $-$0.0013 & 28.8 & $-$695  & $-$5.0 & $-$17.6 & $-$0.3  \\
   \noalign{\smallskip}\hline\noalign{\smallskip}
   $^{139}$La$^{17}$O  & Exp. \cite{knight:jcp1999} (NM) & -0.003(2) & --- & 3751(5) & 29(4) & Abs.val.$<$10 & --- \\
   & Exp. \cite{childs:jms1986} (GP) & --- & 66.1972(5) & 3631.9(1) & 31.472(1) & --- & --- \\
   & B3LYP-U & $-$0.0037 & 73.3 & 3700 & 27.6 & $-$12.0 & $-$0.3 \\
   & B3LYP-R & $-$0.0046 & 91.3 & 3460 & 16.6 & $-$12.5 & $-$0.6 \\
   & PBE0-U & $-$0.0045  & 90.2 & 3470 & 28.7 & $-$8.8 & $-$0.1 \\
   & PBE0-R & $-$0.0054  & 109 & 3220 & 16.3 & $-$11.4& $-$0.5 \\
   \noalign{\smallskip}\hline\noalign{\smallskip}
   $^{67}$Zn$^{1}$H & Exp. \cite{mckinley:jpca2000} (NM) & $-$0.0182(3) & $7.2(1)\times10^3$ (CA) & 630(1) & 15(1) & 503(1) & $-$1(1) \\
   & B3LYP-U & $-$0.0206 & $8.24\times10^3$ & 576 & 22.4 & 567 & $-$0.2  \\
   & B3LYP-R & $-$0.0244 & $9.79\times10^3$ & 616 & 23.8 & 382 & 1.4  \\
   & PBE0-U & $-$0.0201 & $8.05\times10^3$ & 582 & 21.5 & 490 & $-$0.5  \\
   & PBE0-R & $-$0.0240 & $9.60\times10^3$ & 606 & 23.0 & 348 & 1.4 \\
   \noalign{\smallskip}\hline\noalign{\smallskip}
   $^{67}$Zn$^{19}$F& Exp. \cite{knight:jmr1978} (NM) & $-$0.006(1) & $1.3(2)\times10^2$ (CA) & --- & --- & 319(2) & 177(2) \\
   & B3LYP-U & $-$0.0068 & $1.48\times10^2$ & 1230 & 13.4 & 305 & 252  \\
   & B3LYP-R & $-$0.0073 & $1.59\times10^2$ &  1160 & 15.4 & 266 & 210  \\
   & PBE0-U &  $-$0.0071 & $1.55\times10^2$ &  1230 & 12.7 & 280 & 225 \\
   & PBE0-R &  $-$0.0073 & $1.60\times10^2$ & 1140 & 14.7 & 259 & 190  \\
    \noalign{\smallskip}\hline\noalign{\smallskip}
   $^{111}$Cd$^{19}$F & Exp. \cite{knight:jmr1978} (NM) & $-$0.017(2) & $4.8(6)\times10^2$ (CA) & --- & --- & 266(3) & 202(2) \\
   & B3LYP-U & $-$0.0271 & $7.60\times10^2$ & $-$3590 & $-$251 & 632 & 274  \\
   & B3LYP-R & $-$0.0314 & $8.79\times10^2$ & $-$3600 & $-$255 & 567 & 229   \\
   & PBE0-U & $-$0.0278 & $7.78\times10^2$ & $-$3670 & $-$240 & 582 & 252   \\
   & PBE0-R & $-$0.0320 & $8.96\times10^2$ & $-$3630 & $-$246 & 536 & 210  \\
   \noalign{\smallskip}\hline\noalign{\smallskip}
   $^{67}$Zn$^{107}$Ag & Exp. \cite{kasai:jpc1978} (AM) & $-$0.0118(2) & $39(1)$ (CA) & --- & --- & $-$1324(3)$^{*}$ & 0(1) \\
   (optimized)& B3LYP-U & $-$0.0131 & 43.0 & 306 & 6.1 & $-$1390 & 0.6 \\
   & B3LYP-R & $-$0.0158 & 52.0 & 306 & 6.9 & $-$1250 & $-$0.6  \\
   & PBE0-U & $-$0.0133 & 45.1 & 301 & 6.4 & $-$1340 & 1.2 \\
   & PBE0-R & $-$0.0175 & 59.5 & 308 & 7.5 & $-$1190 & $-$0.5  \\
  \noalign{\smallskip}\hline\noalign{\smallskip}
   $^{105}$Pd$^{1}$H & Exp. \cite{knight:jcp1990} (AM) & 0.291(1) & $-1.252(4)\times10^5$ (CA) & $-$823(4) & $-$22(3) & --- & --- \\
   & Exp. \cite{knight:jcp1990} (NM) & 0.291(1) & $-1.252(4)\times10^5$ (CA) & $-$857(4) & $-$16(3) & --- & --- \\
   & B3LYP-U & 0.303 &  $-1.30\times10^5$ & $-$835 & $-$13.5 & 93.3 & 8.4 \\
   & B3LYP-R & 0.266 &  $-1.14\times10^5$ & $-$914 & $-$2.4 & 117 & 7.0  \\
   & PBE0-U & 0.285 & $-1.22\times10^5$ & $-$801 & $-$16.9 & 91.9 & 7.5 \\
   & PBE0-R & 0.248 & $-1.07\times10^5$ & $-$889 & $-$4.2  & 125 & 6.6  \\
  \noalign{\smallskip}\hline\noalign{\smallskip}
   $^{111}$Cd$^{1}$H & Exp. \cite{tan:jcp1994} (GP) & $-$0.0567(2) (CA) & $1.811(6)\times10^4$ & $-$3764(26) & $-$122(6) & 558(10) & --- \\
   & Exp. \cite{varberg:jms2004} (GP) & --- & --- & $-$3766.3(15) & $-$143(1) & 549.8(18) & $-$2.4(8) \\
   & B3LYP-U & $-$0.0597 & $1.95\times10^4$ & $-3510$ & $-$160 & 593 & $-$0.6  \\
   & B3LYP-R & $-$0.0735 & $2.40\times10^4$ & $-3920$ & $-$175 & 374 & 0.9  \\
   & PBE0-U & $-$0.0586 & $1.91\times10^4$ & $-3620$  & $-$155 & 513 & $-$0.9  \\
   & PBE0-R & $-$0.0724 & $2.36\times10^4$ & $-3950$  & $-$171 & 341 & 0.9  \\
  \noalign{\smallskip}\hline\noalign{\smallskip}
  $^{111}$Cd$^{107}$Ag & Exp. \cite{kasai:jpc1978} (AM) & $-$0.0312(2) &68.9(4) & $-$2053(3)$^{*}$ & $-$63(3)$^{*}$ & $-$1327(3)$^{*}$ & 0(1) \\
   & B3LYP-U & $-$0.0339 & 74.9 & $-$1930 & $-$47.7 & $-$1370 & $-$0.5 \\
   & B3LYP-R & $-$0.0400&  88.4 & $-$2010 &  $-$55.4 & $-$1210 & $-$0.6  \\
   & PBE0-U & $-$0.0355 & 80.4 & $-$1910 &  $-$50.5 & $-$1330 & 1.0 \\
   & PBE0-R & $-$0.0442 &  100 & $-$2050 &  $-$60.6 & $-$1150 & $-$0.5  \\
  \noalign{\smallskip}\hline\noalign{\smallskip}
   $^7$Li$^{40}$Ca & Exp. \cite{ivanova:jcp2011} (GP) & $-$0.0068(1) (CA) & 103(2) & --- & --- & --- & ---  \\
   & B3LYP-U & $-$0.0094 & 141 &  310 &  0.0 &  $-$95.0 & $-$5.2   \\
   & B3LYP-R & $-$0.0119 & 179 & 218 & 0.2 & $-$107 & $-$4.6 \\
   & PBE0-U & $-$0.0090 &  134 & 260 & $-$0.3& $-$85.1 & $-$4.9 \\
   & PBE0-R & $-$0.0123 & 184 & 190 & 0.2 & $-$104.4 & $-$4.5 \\
   \noalign{\smallskip}\hline\noalign{\smallskip}
   $^7$Li$^{138}$Ba & Exp. \cite{dincan:jcp1994} (GP) & $-$0.1205(1) (CA) & 1384.5(9) & --- & --- & --- & ---  \\
   & B3LYP-U &  0.854 & -9820 & 172 &  $-$25.7 & 1010 & $-$300   \\
   & B3LYP-R & $-$0.129 &  1480 & 162 &  0.3 &  806 &   28.1  \\
   & PBE0-U & $-$0.086  &  983  &  112 & 0.3 &  836 &   14.0   \\
   & PBE0-R & $-$0.134  & 1540  & 139 &  0.2 &  792 &   28.7 \\
   \noalign{\smallskip}\hline\noalign{\smallskip}
   $^{40}$Ca$^{19}$F & Exp. \cite{childs:jms1981} (GP) & $-$0.00193(1) (CA) & 39.49793(2)  & --- & --- &   122.025(1) & 13.549(1)  \\
   & B3LYP-U & $-$0.00195 & 39.8 & --- & --- & 126 & 8.2  \\
   & B3LYP-R & $-$0.00180 & 37.2 & --- & --- & 127 & 8.0 \\
   & PBE0-U & $-$0.02090 & 43.2 & --- & --- & 102 & 10.0   \\
   & PBE0-R & $-$0.00184 & 38.0 & --- & --- & 112 & 7.4 \\
   \noalign{\smallskip}\hline\noalign{\smallskip}
   $^{88}$Sr$^{19}$F & Exp. \cite{childs:jms1981b} (GP) & -0.00495(1) (CA) & 74.79485(10) & --- & --- & 107.1724(10) & 10.089(10)  \\
   & B3LYP-U & $-$0.00431 & 65.1 & ---  & --- &  114 & 7.0 \\
   & B3LYP-R & $-$0.00463 & 69.9 & --- & --- & 112 & 6.8 \\
   & PBE0-U & $-$0.00469 & 65.1 & --- & --- & 90.8 &  8.1   \\
   & PBE0-R & $-$0.00485 & 73.2 & --- & --- & 98.5 & 6.1 \\
   \noalign{\smallskip}\hline\noalign{\smallskip}
\end{longtable*}

\begin{acknowledgments}
This work was supported by the U.K. Engineering and Physical Sciences Research
Council (EPSRC) Grants No.\ EP/H003363/1, EP/I012044/1, EP/P008275/1 and
EP/P01058X/1. JA acknowledges funding by the Spanish Ministry of Science and
Innovation Grants No.\ CTQ2012-37404-C02, CTQ2015-65033-P, and Consolider
Ingenio 2010 CSD2009-00038.
\end{acknowledgments}

%

\end{document}